\documentclass[cits]{PoS}

\def\apj{ApJ}%
\def\apjl{ApJ}%
\def\aap{A\&A}%
\def\nat{Nature}%

\title{Pre-supernova models at very low metallicity}

\ShortTitle{Pre-SN models at very low Z}

\author{\speaker{Raphael Hirschi}\\
        Dept. of physics and Astronomy, University of Basel,           
	      Klingelbergstr. 82, CH-4056 Basel\\
        E-mail: \email{Raphael.Hirschi@unibas.ch}}

\abstract{
A series of fast rotating models at very low metallicity ($Z=10^{-8}$) was computed in order to 
explain the surface abundances
observed at the surface of CEMP
stars, in particular for nitrogen. The
main results are the following:
\begin{list}{-}{}
\item Strong mixing occurs during He--burning and leads to important primary nitrogen production.
\item Important mass loss takes place in the RSG stage for the most massive models.
The 85 $M_\odot$ model loses about three quarter of its initial mass, becomes a 
WO star and could produce a GRB.
\item The CNO elements of HE1327-2326 could have 
been produced in massive rotating stars and ejected by their stellar winds.
\end{list}
}

\FullConference{International Symposium on Nuclear Astrophysics --- Nuclei in the Cosmos --- IX\\
		 June 25-30 2006\\
		 CERN, Geneva, Switzerland}

\begin{document}

\section{Introduction}
Precise measurements
of surface abundances of extremely metal poor (EMP) stars have recently 
been obtained \cite{FS5,FS6,IER04}. 
These provide new constraints for 
stellar evolution models (see \cite{CMB05,Fr04,Pr05}).
The most striking constraint is the need for primary $^{14}$N production in very low
metallicity massive stars. 
About one quarter of EMP stars are carbon rich (C-rich EMP, CEMP stars).
Ryan et al \cite{RANB05} propose a classification for these CEMP stars. They find two
categories: about three quarter are main s--process enriched (Ba-rich) 
CEMP stars and one quarter are enriched with a weak component of
s--process (Ba-normal). The two most metal poor stars known to date, 
HE1327-2326 \cite{Fr05,Ao06} and HE 0107-5240 \cite{Ch04} are both
CEMP stars. These stars are thought to have been enriched by only one
to several stars and the yields of the models can therefore be compared to
their observed abundances without the filter of a galactic chemical
evolution model.

The evolution of very low metallicity or metal free stars is not a new
subject (see for example \cite{CC83,EE83}). The observations
cited above have however greatly increased the interest in very metal
poor stars. There are many recent works studying the evolution of metal
free (or almost) massive \cite{HW02,LC05,UN05,MEM06}, intermediate mass
\cite{SAMFI04} and low
mass \cite{W04} stars. 
In this work pre-supernova
evolution models of rotating single stars were computed 
at a metallicity, $Z=10^{-8}$ to study the impact of
rotation in the evolution of very low metallicity stars.

\section{Computer model \& calculation}
The computer model used here is the same as the one described in
 \cite{psn04}. Convective stability is determined 
by the Schwarzschild criterion. Overshooting is only considered for 
H-- and He--burning cores with an overshooting parameter, 
$\alpha_{\rm{over}}$, of 0.1 H$_{\rm{P}}$. 
Models were computed at Z=10$^{-8}$ with initial masses of 20,
40, 60 and 85 $M_\odot$ and initial rotational velocities of 600, 700, 800 and
800\,km\,s$^{-1}$ respectively. 
At first sight, these velocities appear to be 
very high but their
corresponding total angular momentum is similar to the one contained in solar 
metallicity stars with rotational velocities of 300\,km\,s$^{-1}$.
The evolution of the models was followed until core Si--burning except for 
the 60 $M_\odot$, which was followed until neon burning. 
The yields of these models were calculated in
the same way as in \cite{ywr05}. The (pre--)SN yields were calculated 
without taking into account the explosive nucleosynthesis. 
We therefore only present yields of light elements,
 which are not significantly affected by the subsequent
evolution. 
\begin{figure}
\includegraphics[width=.5\textwidth]{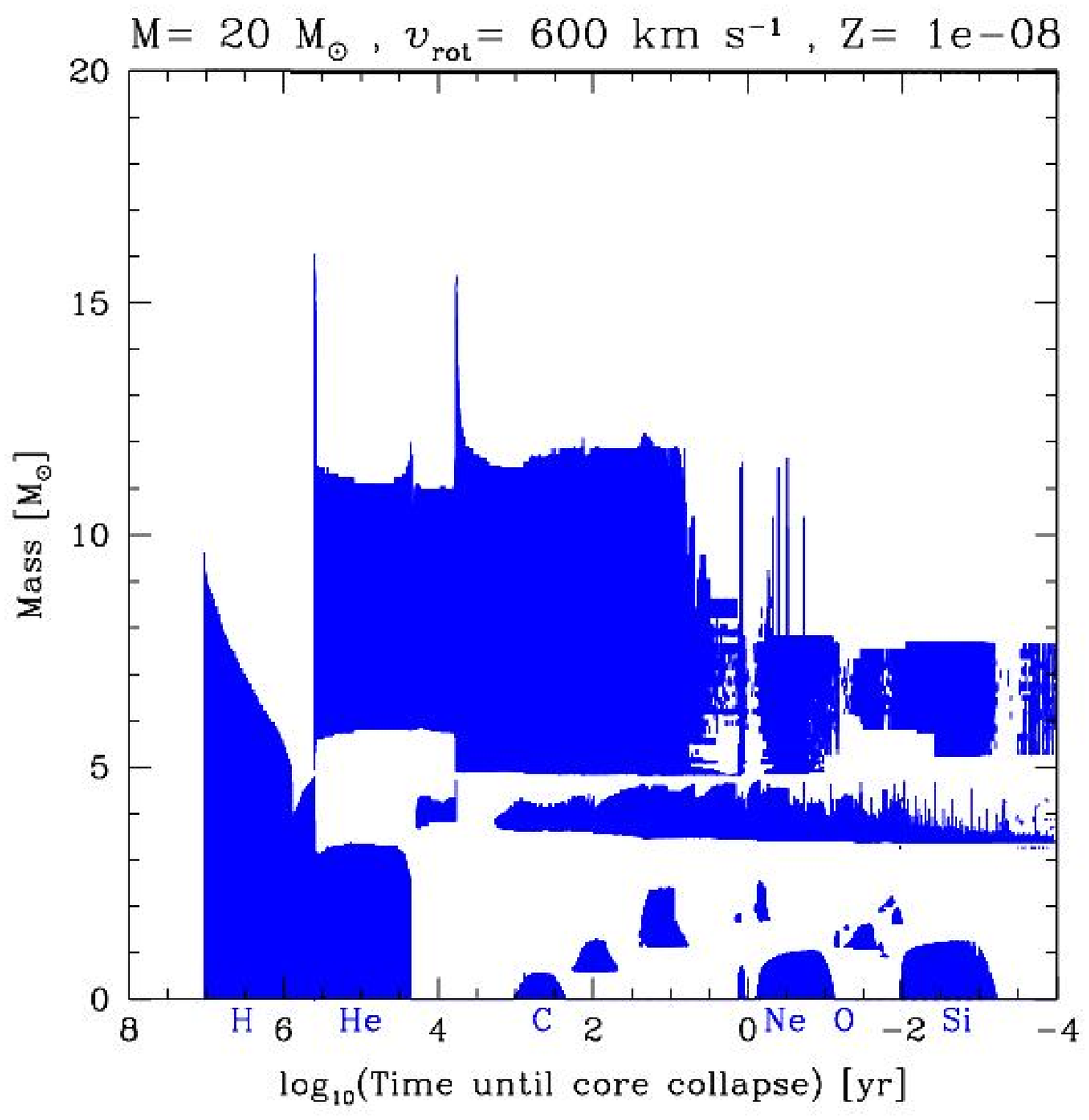}\includegraphics[width=.5\textwidth]{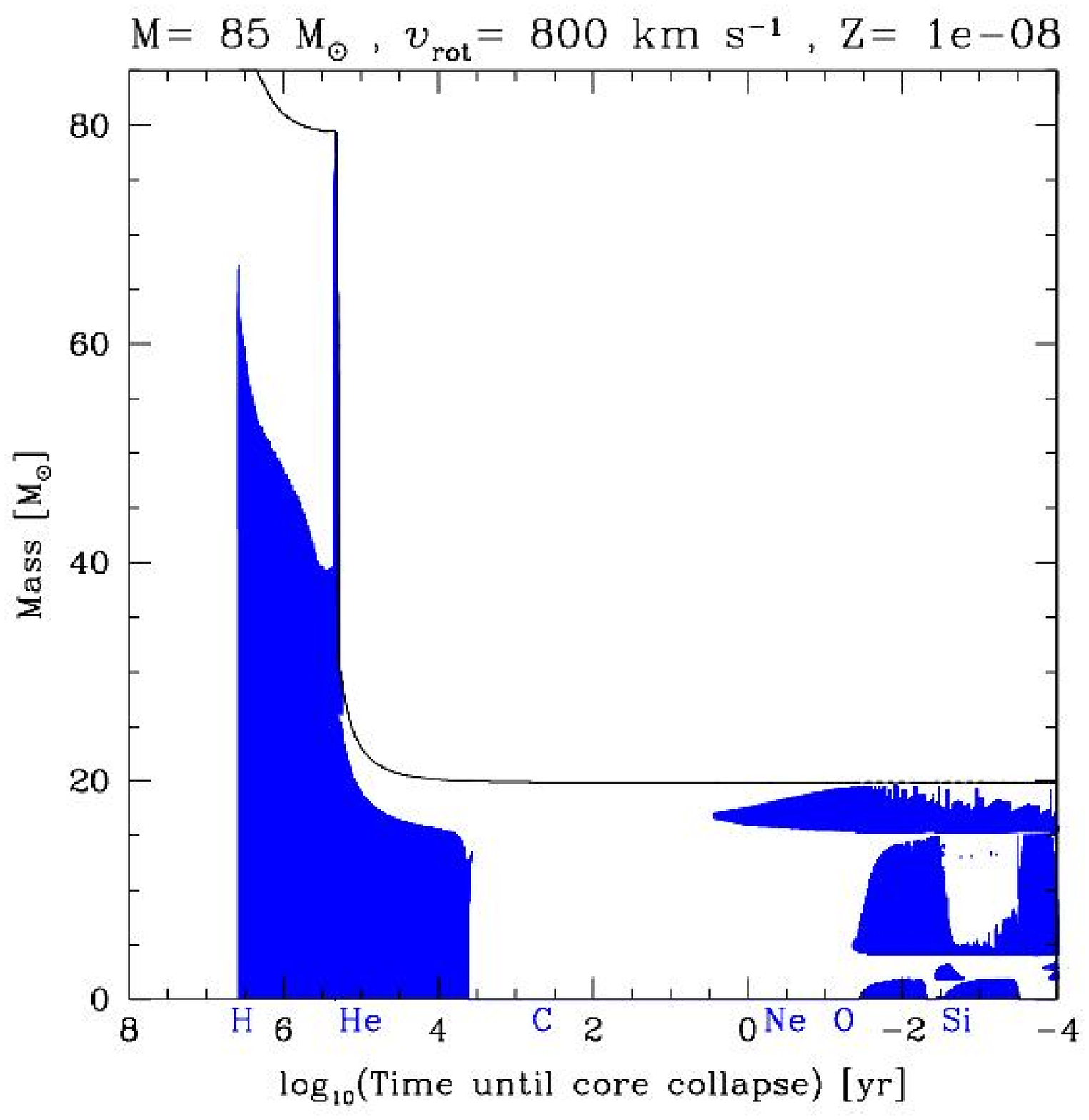}
\caption{Stellar structure (Kippenhahn) diagrams, which show the
evolution of the structure as a function of the time left until the core
collapse for the 20 $M_\odot$ ({\it left}) and 85 $M_\odot$ ({\it right}). 
The coloured zone correspond to the convective zones and the symbols of
the burning stages are given below the time axis.
}
\label{fig1}
\end{figure}
\begin{figure}
\begin{center}
\includegraphics[width=.8\textwidth]{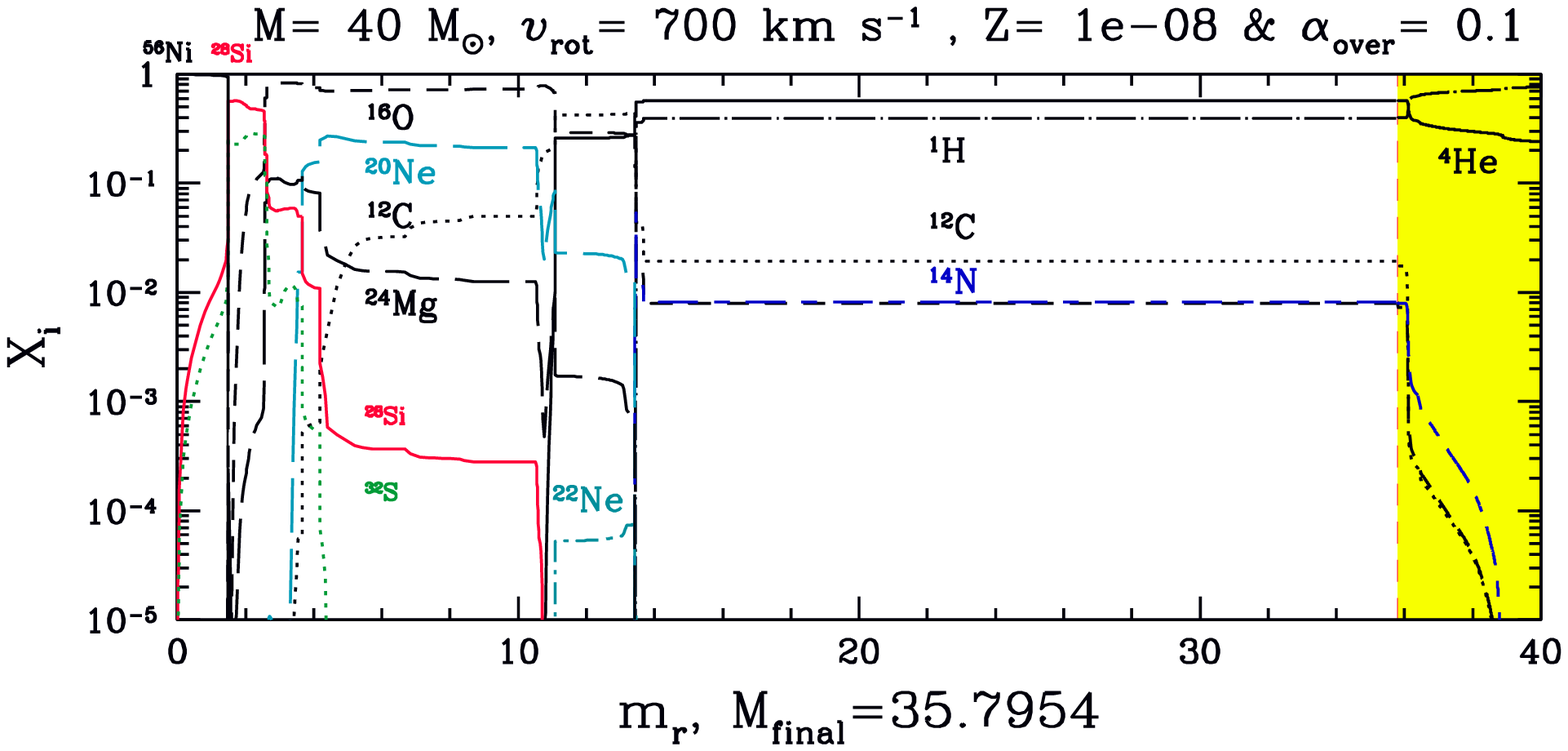}
\includegraphics[width=.8\textwidth]{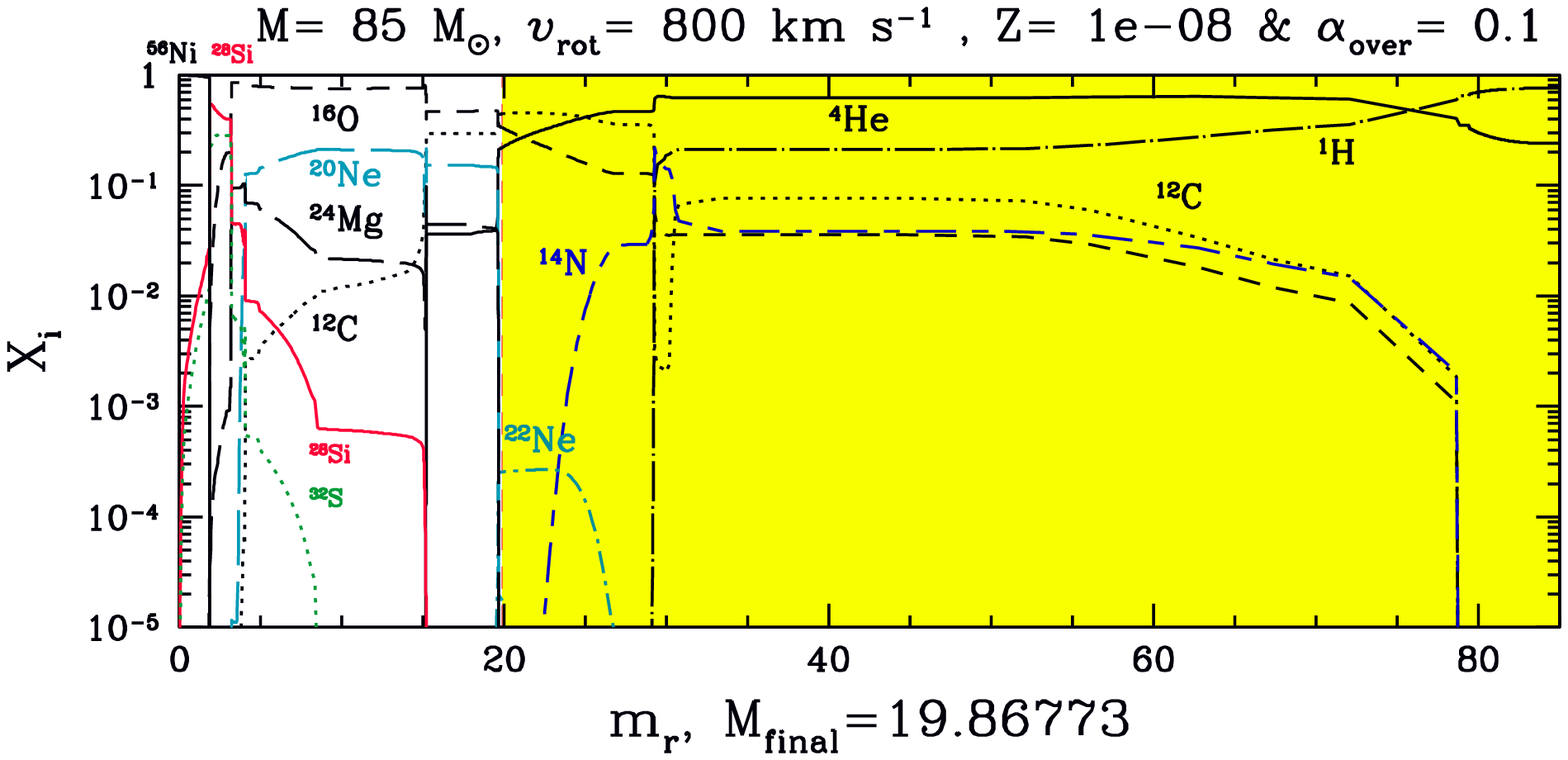}
\end{center}
\caption{
Abundance profiles for the 40 $M_\odot$ ({\it top}) and 85 $M_\odot$ ({\it 
bottom}) models. The pre--SN profiles and wind  (yellow shaded
area) profiles are
separated by a red dashed line located at the pre--SN total mass
($M_{\rm final}$), given below each plot.
}
\label{fig2}
\end{figure}
\section{Evolution of the structure}
The evolution of the structure of the models is shown in Fig. 1.
A strong mixing takes place during He--burning.
Primary carbon and oxygen are
mixed outside of the convective core into the H--burning shell. Once the
enrichment is strong enough, the H--burning shell is boosted. 
The H--burning shell then becomes convective.
In response to the shell boost, the core
expands and the convective core mass decreases.
At the end of He--burning, the CO core is smaller in mass than in non--rotating
models (see \cite{H06}).
The yield of $^{16}$O being closely
correlated with the mass of the CO core, it is therefore
reduced due to the strong mixing. 
At the same time the carbon yield is slightly increased.
It is interesting to note that the shell H--burning boost occurs in all
the different initial mass models at $Z=10^{-8}$.
This means that the strong mixing could
be an explanation for the possible high [C/O] ratio observed in the most
metal poor halo stars (see Fig. 14 in \cite{FS6}).

The 85 $M_\odot$ model becomes a WO type (see Fig. 2) WR star. SNe of type Ib,c are
therefore expected to ensue from the death of single massive stars at
very low metallicities. The core of the 85 $M_\odot$ model retains enough
angular momentum to produce a GRB via the collapsar model 
(see \cite{YL05,WH06,grb05} for more details on GRB progenitors). 

\section{Wind and pre--SN composition and comparison with CEMP stars}

\begin{figure}
\includegraphics[width=.5\textwidth]{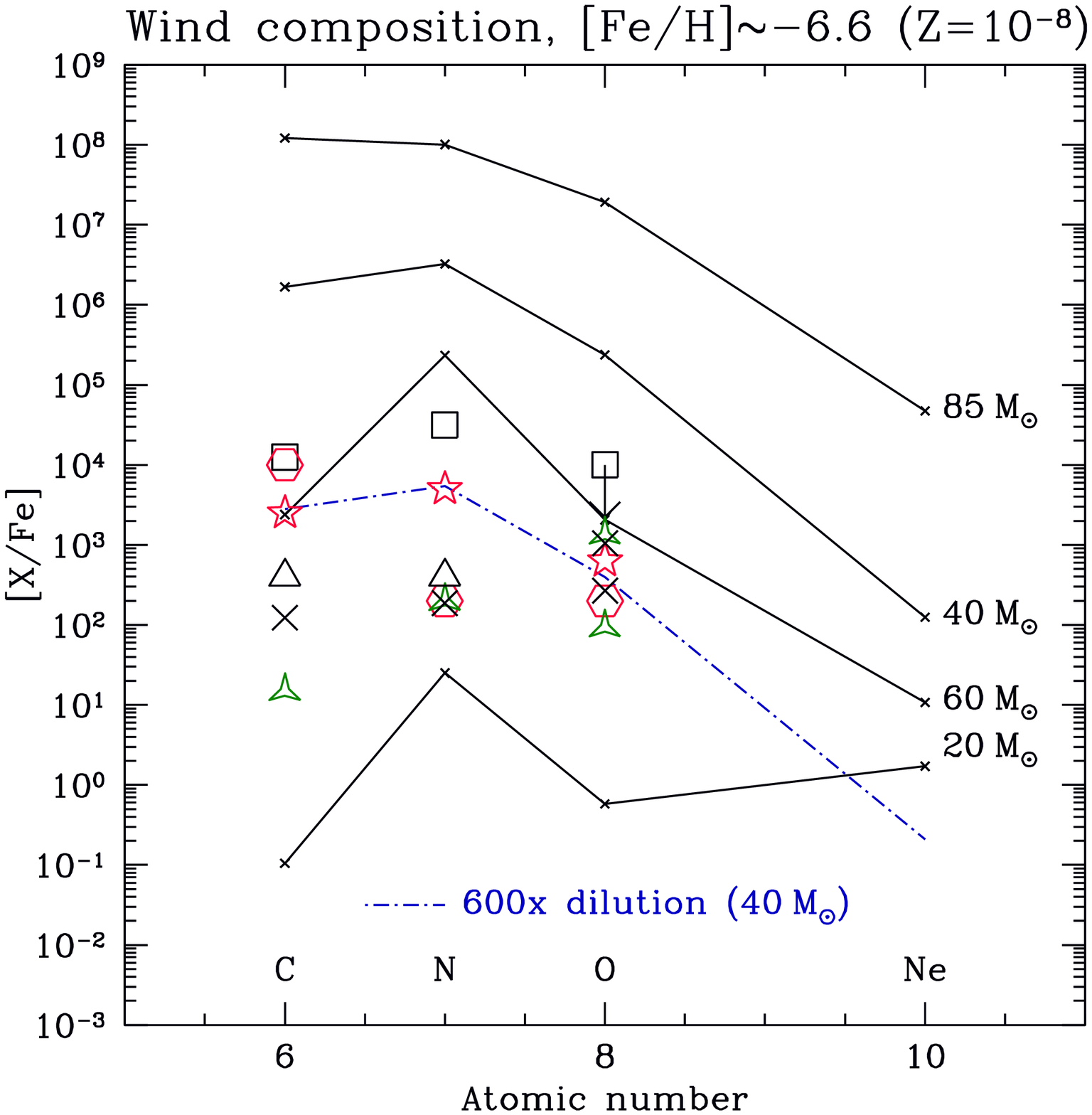}
\includegraphics[width=.5\textwidth]{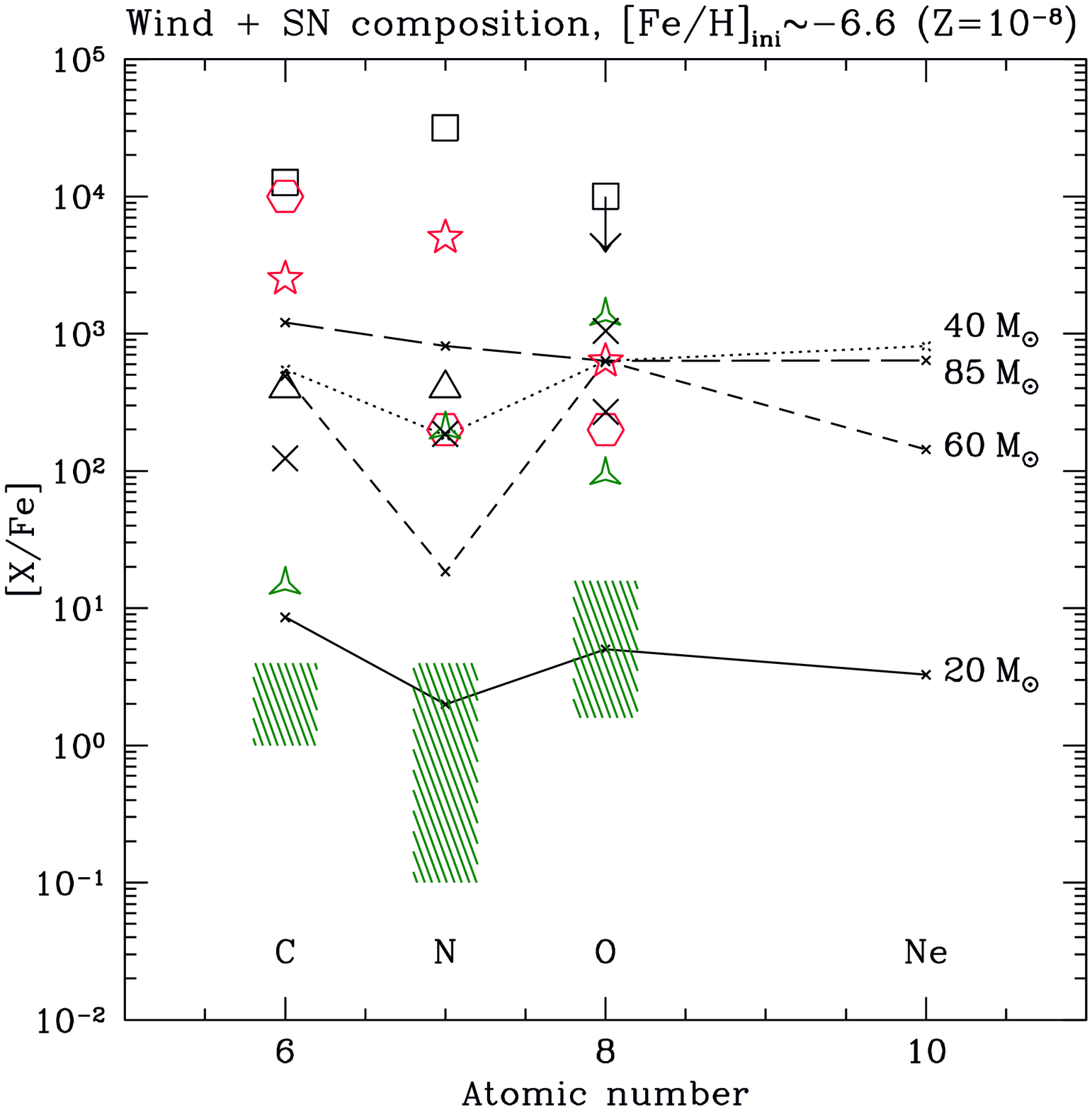}
\caption{Composition in [X/Fe] of the stellar wind ({\it left}) and the
sum of the wind and SN ejecta ({\it right}) for the $Z=10^{-8}$ models.
The lines represent predictions from the models.
The {\it empty triangles} \cite{PC05}, 
and {\it squares} \cite{Fr05}, [Fe/H]$\simeq -5.4$
correspond to non-evolved CEMP stars.
The new (3D/NLTE corrected) estimates for HE1327-2326 from \cite{Fr06} are
represented by the {\it red stars}. 
On the right, wind+SN ejecta is compared to the normal EMP
stars \cite{FS5,FS6} for the 20 $M_\odot$ model (green hatched area) 
and again to the CEMP stars for the more
massive models. For this purpose, the value [O/Fe] is chosen
to fall in the middle of the observed range (20 $M_\odot$: [O/Fe]=0.7 and
$M> 20\,M_\odot$: [O/Fe]=2.8).
The other symbols correspond to the 
abundances measured at the surface of giant CEMP stars.
}
\label{fig3}
\end{figure}

The CNO abundances of HE1327-2326 could be explained with the following scenario. 
A first generation of stars (PopIII) 
pollutes the interstellar medium to very low metallicities
([Fe/H]$\sim$-6). Then a PopII.5 star like the 
40 $M_\odot$ model calculated here
pollutes (mainly through its wind) again the interstellar medium out of
 which HE1327-2326 forms.
In this scenario, 
the CNO abundances are well reproduced, in particular that of
nitrogen, which according the new values for a subgiant from \cite{Fr06}
is 0.9 dex higher in [X/Fe] than oxygen. 
This is shown in Fig. 3 where the new abundances are
represented by the red stars and the best fit is 
obtained by diluting the composition of the wind of the 40 $M_\odot$
model by a factor 600. On the right side, one sees that when the SN
contribution is added, the [X/Fe] ratio is usually lower for nitrogen
than for oxygen.


\end{document}